# Asymmetric Magnetic Domain-Wall Motion by the Dzyaloshinskii-Moriya Interaction

Soong-Geun Je,[1] Duck-Ho Kim,[1] Sang-Cheol Yoo,[1,2] Byoung-Chul Min,[2] Kyung-Jin Lee,[3,4] and Sug-Bong Choe[1*]

[1]*Center for Subwavelength Optics and Department of Physics, Seoul National University, Seoul 151-742, Republic of Korea*

[2]*Center for Spintronics Research, Korea Institute of Science and Technology, Seoul 136-791, Republic of Korea*

[3]*Department of Materials Science and Engineering, Korea University, Seoul 136-701, Republic of Korea*

[4] *KU-KIST Graduate School of Converging Science and Technology, Korea University, Seoul 136-713, Korea*

*Correspondence to: sugbong@snu.ac.kr

We demonstrate here that ultrathin ferromagnetic Pt/Co/Pt films with perpendicular magnetic anisotropy exhibit a sizeable Dzyaloshinskii-Moriya interaction (DMI) effect. Such a DMI effect modifies the domain-wall (DW) energy density and consequently, results in an asymmetric DW expansion driven by an out-of-plane magnetic field under an in-plane magnetic field bias. From an analysis of the asymmetry, the DMI effect is estimated to be strong enough for the DW to remain in the Néel-type configuration in contrast to the general expectations of these materials. Our findings emphasize the critical role of the DMI effect on the DW dynamics as the underlying physics of the asymmetries that are often observed in spin-transfer-related phenomena.

The in-plane magnetic field that is applied to perpendicularly magnetized films has drawn great attention because of its crucial role in current-induced magnetization dynamics [1-4]. Recent experiments have shown that the directions of both current-induced magnetization switching [1, 3] and the domain-wall (DW) motion [4] are governed by the polarity of the in-plane magnetic field, even when the magnetic field is much weaker than the perpendicular anisotropy field. Such phenomena are attributed to the in-plane-field-induced magnetization tilt, which initiates the additional torques that are caused by either the spin Hall effect [2-8] or the Rashba [1, 8-10] effect from the current injection. These findings have promptly motivated numerous studies because of the technological interest on an alternative method of magnetization control [1-3, 11] and the academic debate on the hierarchy between the spin Hall effect and the Rashba effect [12-15].

Interestingly, even without the current-induced spin Hall effect or Rashba effect, an in-plane magnetic field is found to affect the purely magnetic-field-driven magnetization dynamics, as demonstrated in Fig. 1 for the DW motions in a ferromagnetic Pt/Co/Pt film. Each image in the figure is obtained by adding several sequential images; thus, each image shows several DWs in motion simultaneously. The circular domain is magnetized along the $+z$ direction as shown by the brighter contrast.

The figure clearly shows that when a circular domain expands under an out-of-plane magnetic field, the center of the circular domain shifts along the direction of the in-plane magnetic field (Fig. 1(b)), which contrasts the concentric expansion without an in-plane magnetic field (Fig. 1(a)). This observation appears strange at first because the in-plane magnetic field does not generate any energy gradient for the center of the domains to move.

One of the possible origins of this phenomenon might be the symmetry breaking related to the anti-symmetric exchange interaction—the so-called Dzyaloshinskii-Moriya interaction (DMI) [16, 17]—which prefers a helical magnetic order and consequently forms the Néel-type DW [6, 18, 19] instead of the Bloch-type DW in perpendicularly magnetized thin films. This DMI was originally studied in chiral magnets [20-24], but a sizeable DMI was recently observed in ferromagnetic thin films with an asymmetric layer structure [18, 19, 25, 26]. For a circular domain, the DMI induces an effective magnetic field on the DW in the radial direction and maintains the rotational symmetry with respect to the axis parallel to the out-of-plane magnetic field. Therefore, it is natural to observe an isotropic DW expansion as shown in Fig. 1(a). However, with the application of an in-plane magnetic field, such rotational symmetry is broken and thus, it becomes possible for the DW to show anisotropic expansion as shown in Fig. 1(b).

To examine whether this scenario actually occurs, we first examine the DW motion that is driven by an out-of-plane magnetic field $H_z$ under an in-plane magnetic field bias $H_x$. Figure 2 shows the two-dimensional contour map $V(H_x,H_z)$ of the DW speed $V$ as a function of $H_x$ and $H_z$. Here, $V$ is measured by detecting the DW displacement at the rightmost place of the circular domain (indicated by the blue box in Fig. 1(b)), where the DW lies normal to $H_x$ and displacement occurs along the $+x$ axis [27]. Because the color in the map corresponds to the magnitude of $V$ with the scale shown on the right, each color traces an equi-speed contour [28]. Several equi-speed contours are highlighted by the circular symbols, of which the position $(H_x,H_z)$ indicates the value of $H_z$ for each $H_x$ on each equi-speed contour.

The contour map clearly shows that all equi-speed contours exhibit an inversion symmetry with respect to the axis $H_x=H_0$, where $H_0$ is a constant. The symmetry axis is shown with the vertical purple line on the map. For better visualization, the cross symbols are added on the map at positions $(H_0-H_x,H_z)$, which are the mirrored positions of the circular symbols at $(H_x,H_z)$. It is clear from the figure that the two types of symbols are overlapped onto the same curves to manifest the inversion symmetry with respect to $H_0$. The best value of $H_0$ is -26.5±0.5 mT. Such non-zero offset $H_0$ of the symmetry axis can be attributed to the DMI effect because DMI induces an effective magnetic field $H_{DMI}$ along the *x* axis in this geometry; thus, the DW experiences the resultant magnetic field $H_x+H_{DMI}$. For this case, the experimental value $H_0$ can be considered a direct measure of $H_{DMI}$, and the negative sign indicates that the direction of $H_{DMI}$ inside the DW is parallel to the +*x* axis, which points from the domain that is magnetized along the +*z* axis to the domain that is magnetized along the -*z* axis.

Next, we consider the possible effects of DMI on the shape of the equi-speed contours. Figure 3(a) plots the value $H_z$ of the circular symbols in the map, normalized by the average $<H_z>$ for each equi-speed contour with respect to $H_x$. Interestingly, the results show that all of the data are collapsed onto a single curve. This observation indicates that all normalized values $H_z/<H_z>$ follow a unique function, which is denoted as $f(H_x)$ hereafter. In the calculation of $<H_z>$ for each equi-speed contour, a wider averaging range of $H_x$ is better to reduce the experimental noise, but the same relation is essentially observed regardless of the range. Thus, the average $<H_z>$ can be replaced by a single value of $H_z$ when the selected range is as narrow as possible. By choosing a narrow range adjacent to $H_x=0$, the observed relation can be written in the form of a separation of variables as follows

$$H_z = H_z^*(V) \cdot f(H_x), \tag{1}$$

where $H_z^*(V)$ denotes the value of $H_z$ at $H_x=0$ on the contour with speed $V$. The present definition of $H_z^*$ leads to the relation $f(0)=1$.

For the conventional field-driven DW motion with $H_x=0$, it is well known that the DW motion follows the DW creep scaling law [29-31] in the present experimental range of $H_z$. In the creep law, the DW speed $V$ is given by $V=V_0\exp(-\alpha H_z^{-\mu})$, where $V_0$ is the characteristic speed, and $\alpha$ is a scaling constant. The creep scaling exponent $\mu$ is 1/4 [29, 32]. This conventional law can be modified to the relation $H_z^*(V)=[\ln(V_0/V)/\alpha]^{-1/\mu}$ based on the definition of $H_z^*$. By adopting Eq. (1) into this relation, one finds that the DW speed with $H_x \neq 0$ also follows an identical creep scaling law as given by

$$V = V_0 \exp\left(-\alpha^* H_z^{-\mu}\right), \tag{2}$$

except $\alpha$ is replaced by $\alpha^*$, which is defined as

$$\alpha^*(H_x) \equiv \alpha f^{\mu}(H_x). \tag{3}$$

The experimentally determined values of $\alpha^*(H_x)$ from the best fit are plotted in Fig. 3(b). To check the validity of this approach, the inset shows $V$ vs. $\alpha^* H_z^{-\mu}$ for all of the experimental data. It is clear from the figure that all data are collapsed onto a single curve, confirming that all the data follows the same scaling law given by Eq. (2). The best fitting value of $V_0$ is found to be $(8.4\pm0.4)\times10^3$ m/s.

The scaling constant $\alpha$ is originally defined as $U_C H_{crit}^{\mu}/k_B T$ in the DW creep theory [29], where $U_C$ is the energy constant, $H_{crit}$ is the critical magnetic field, and $k_B T$ denotes the thermal fluctuation energy. According to Ref. [32] (Supplementary Information V), $U_C$ and $H_{crit}$ are defined as $U_C \equiv [\mu u_C/2(\mu+1)\xi]^{\mu} \sigma_{DW} t_f u_C^2/(1+\mu)L_C$ and $H_{crit} \equiv \sigma_{DW} \xi/M_S L_C^2$, respectively, where $\xi$ is the correlation length of the disorder potential, $u_C$ is the roughness of the DW segment with length $L_C$, and $L_C$ is the Larkin length that is the characteristic length of rigid microscopic DW segments. $M_S$, $t_f$, and $\sigma_{DW}$ are the saturation magnetization, the film thickness, and the DW energy density per unit area, respectively. Applying the relation $L_C = (\sigma_{DW}^2 t_f^2 \xi^2/\gamma)^{1/3}$ [33] with the pinning strength $\gamma$ of the disorder, $\alpha$ can be written as a function of $\mu$, $\gamma$, $u_C$, $\xi$, $t_f$, $k_B T$, $M_S$, and $\sigma_{DW}$. Because all other parameters except $\sigma_{DW}$ do not depend on a magnetic field, the field dependence of $\alpha$ can be solely attributed to the field dependence of $\sigma_{DW}$, i.e., $\alpha(H_x) \propto [\sigma_{DW}(H_x)]^{1/4}$ or

$$\alpha(H_x)=\alpha(0)\cdot[\sigma_{DW}(H_x)/\sigma_{DW}(0)]^{1/4}. \tag{4}$$

Note that Eq. (4) is identical to the empirical Eq. (3) by equating $f(H_x)=\sigma_{DW}(H_x)/\sigma_{DW}(0)$. It is therefore possible to conclude that the experimentally-observed $H_x$-dependence of the DW speed and consequently, the shape of the equi-speed contour are attributed to the variation of the DW energy density with respect to $H_x$.

Recent studies [6, 18] on the DMI effect on DWs have proposed that $\sigma_{DW}$ is given by

$$\sigma_{DW}(H_x,\psi)=\sigma_0+2K_D\lambda\cos^2\psi-\pi\lambda M_S(H_x+H_{DMI})\cos\psi, \tag{5}$$

where $\sigma_0$ is the Bloch-type DW energy density, $K_D$ is the DW-anisotropy energy density [34], and $\lambda$ is the DW width. The angle $\psi$ of the magnetization direction inside the DW is defined as the azimuthal angle from the $+x$ axis. From the minimization condition $\partial\sigma_{DW}/\partial\psi=0$, the equilibrium angle $\psi_{eq}$ can be obtained as $\cos\psi_{eq}=\pi M_S(H_x+H_{DMI})/4K_D$. Then, the DW energy with the equilibrium angle is given by

$$\sigma_{DW}(H_x) = \begin{cases} \sigma_0 - \dfrac{\pi^2 \lambda M_S^2}{4K_D}(H_x + H_{DMI})^2 & \text{for } |H_x + H_{DMI}| < \dfrac{4K_D}{\pi M_S} \\ \sigma_0 + 2K_D\lambda - \pi\lambda M_S|H_x + H_{DMI}| & \text{otherwise} \end{cases} \quad (6)$$

where $4K_D/\pi M_S$ is the magnetic field required to saturate $\psi_{eq}$ to 0. The best fit using Eqs. (4) and (6) is plotted with the black solid lines in Figs. 3(a) and (b) and also, the equi-speed contour lines in Fig. 2. The best fitting parameters are found within the range of the typical values known for Pt/Co/Pt films, which are $\sigma_0$=4.7±0.3 mJ/m$^2$ and $K_D$=(1.4±0.1)×10$^4$ J/m$^3$ with $M_S$=1 T [35] and $\lambda$=5 nm. The value of $H_{DMI}$ estimated from Fig. 2(a) is used in the present fitting. The good consistency with the experimental data verifies the role of $H_{DMI}$ on the DW energy density and the shape of the equi-speed contours.

The present theory can be readily extended for a magnetic field with an arbitrary angle by simply inserting the term $-\pi\lambda M_S H_y \sin\psi$ into Eq. (5). The dashed red circles in Fig. 1(b) show the calculation results of the DW shape after the asymmetric expansion using the best fitting parameters determined from Fig. 3(b). The exact match to the shape of the circular image validates the concept of the present theory, which explains the asymmetric DW expansion with respect to $H_x$.

Finally, we examine the helicity of $H_{DMI}$ with respect to the magnetization direction of the neighboring domains. For this examination, another experiment with opposite magnetic polarities is performed, in which a circular domain is magnetized along the $-z$ direction and the outer domain is magnetized along the $+z$ direction. Note that all polarities of the domains in this latter experiment are opposite to those shown in Fig. 1 for the former experiment. To expand the circular domain, a magnetic field is applied along the $-z$ direction. Figure 4 summarizes the results. It is clear from the figure that the latter experiment exhibits essentially an identical behavior, except the opposite sign of $H_{DMI}$ in comparison to the former experiment. This observation implies that $H_{DMI}$ is always pointing from the domain magnetized along the $+z$ direction to the domain magnetized along the $-z$ direction. Therefore, the chirality is maintained identical for both experiments in accordance with the prediction on the basic properties of $H_{DMI}$ [6, 18, 19, 25, 26].

In the latter experiment, if we rotate the observation coordinate by 180° with respect to the $x$ axis, all polarities of the domains and the external magnetic field in the new coordinate become identical to those shown in Fig. 1 for the former experiment. Nevertheless, the signs of $H_{DMI}$ of these two experiments remain opposite to each other. The opposite sign of $H_{DMI}$ is inevitably attributed to the flipping of the sample in the new observation coordinate for the latter experiment; thus, the layer structures in the two experiments have opposite asymmetries. It is therefore natural to understand that the $H_{\text{DMI}}$ values have opposite signs because the DMI effect is caused by the asymmetry of the layer structure [18, 25]. Thus, the present experiments prove the direct relation between the sign of $H_{\text{DMI}}$ and the asymmetry of the layer structure.

However, it is surprising that the magnitude of $H_{\text{DMI}}$ is notably strong, although it is commonly expected [2-8] that the asymmetry is small in the present sample because the magnetic Co layer is sandwiched between identical nonmagnetic Pt layers and because the layer structure (2.5-nm Pt/0.3-nm Co/1.5-nm Pt) is almost symmetric with only a small thickness difference. Note that the measured $H_{\text{DMI}}$ is larger than the DW anisotropy field $4K_{\text{D}}/\pi M_{\text{S}}$ (=22±2 mT). Thus, the DWs in this sample are expected to spontaneously remain in the Néel-type configuration even without any in-plane magnetic field.

Such sizeable $H_{\text{DMI}}$ might be caused by the asymmetric interface formation because the interface that is formed by depositing Co atoms onto a Pt layer is generally different from the interface that is formed by another sequence [36, 37]. Then, such a different interface structure induces the asymmetric interfacial structure. These interfacial effects are expected to be crucial in the present sample because the sample has only approximately 1.5 monolayers of Co atoms. Thus, all Co atoms must be completely influenced by the interfacial structure.

In conclusion, we demonstrate here that the asymmetric expansion observed in purely field-driven DW motion is attributed to the DMI effect. This DMI effect modifies the DW energy density by tilting the magnetization direction inside the DW and consequently affects the DW creeping speed. Our experiment on the asymmetry directly quantifies the DMI-induced effective field, which is found large enough to induce the Néel-type DW in ferromagnetic Pt/Co/Pt films.

This work was supported by the National Research Foundation of Korea (NRF) grant funded by the Korea government (MSIP) (2012-003418, 2008-0061906). SGJ was supported by the National Research Foundation of Korea (NRF) grant funded by the Korea government (NRF-2011-Global Ph.D. Fellowship Program). SCY and BCM were supported by the KIST institutional program and pioneer research center program (2011-0027905) through NRF funded by MEST. KJL was supported by NRF (NRF-2013R1A2A2A01013188).

[27]    For better experimental accuracy, the DW speed at the leftmost end of the circular domain under $+H_x$ is also measured and then, averaged with the DW speed under $-H_x$ measured at the rightmost end. This measurement is done for all the $H_x$ range [-60 mT, +60 mT]. The in-plane magnet is aligned to the film plane within an accuracy of ±0.2° and the possible effect from such small misalignment as well as the ambient magnetic field is included in the $H_z$ calibration.
[28]    J.-C. Lee *et al*., Phys. Rev. Lett. **107**, 067201 (2011).

[29]    S. Lemerle *et al*., Phys. Rev. Lett. **80**, 849 (1998).

[30]    P. Chauve, T. Giamarchi, and P. Le Doussal, Phys. Rev. B **62**, 6241 (2000).

[31]    J. Ryu, S.-B. Choe, and H.-W. Lee, Phys. Rev. B **84**, 075469 (2011).

[32]    K.-J. Kim *et al*., Nature (London) **458**, 740 (2009).

[33]    F. Cayssol, D. Ravelosona, C. Chappert, J. Ferré, and J. P. Jamet, Phys. Rev. Lett. **92**, 107202 (2004).

[34]    S.-W. Jung, W. Kim, T.-D. Lee, K.-J. Lee, and H.-W. Lee, Appl. Phys. Lett. **92**, 202508 (2008).

[35]    P. J. Metaxas *et al*., Phys. Rev. Lett. **99**, 217208 (2007).

[36]    R. Lavrijsen *et al*., Appl. Phys. Lett. **100**, 262408 (2012).
16

17[37]    S. Bandiera, R. C. Sousa, B. Rodmacq, and B. Dieny, IEEE Magn. Lett. **2**, 3000504 (2011).17

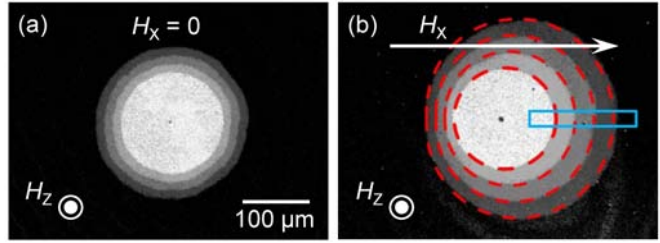

**FIG. 1.** Circular DW expansion driven by an out-of-plane magnetic field $H_z$ (3 mT), (a) without an in-plane magnetic field and (b) with an in-plane magnetic field $H_x$ (50 mT). Each image is obtained by adding four sequential images with a fixed time step (0.4 s), which are captured using a magneto-optical Kerr effect microscope. The white arrow and the symbols indicate the directions of each magnetic field. The blue box in (b) designates where the DW displacement is measured. The dashed red circles in (b) show the calculation results based on Eq. (5) with an extension to arbitrary angles.

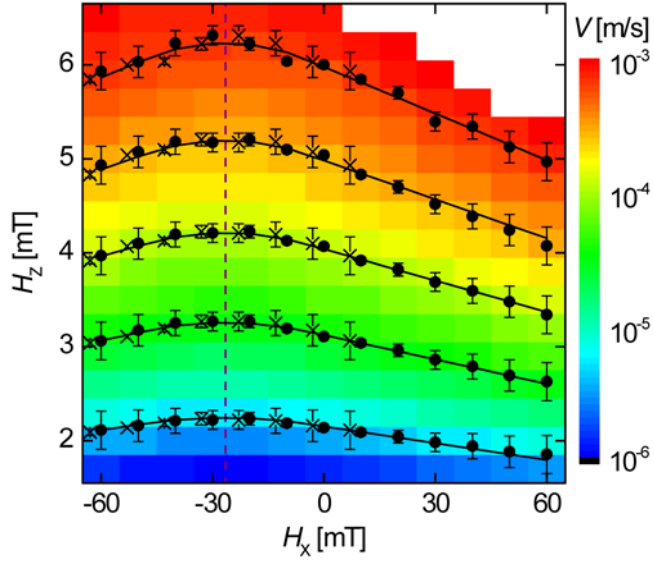

**FIG. 2.** Two-dimensional equi-speed contour map of $V$ as a function of $H_x$ and $H_z$. The color corresponds to the magnitude of $V$ with the scale on the right. The symbols with error bars show the measured positions ($H_x$,$H_z$) on several equi-speed contours. The black solid lines show the best fit using Eq. (2). The purple line indicates the symmetric axis $H_x=H_0$ for inversion.

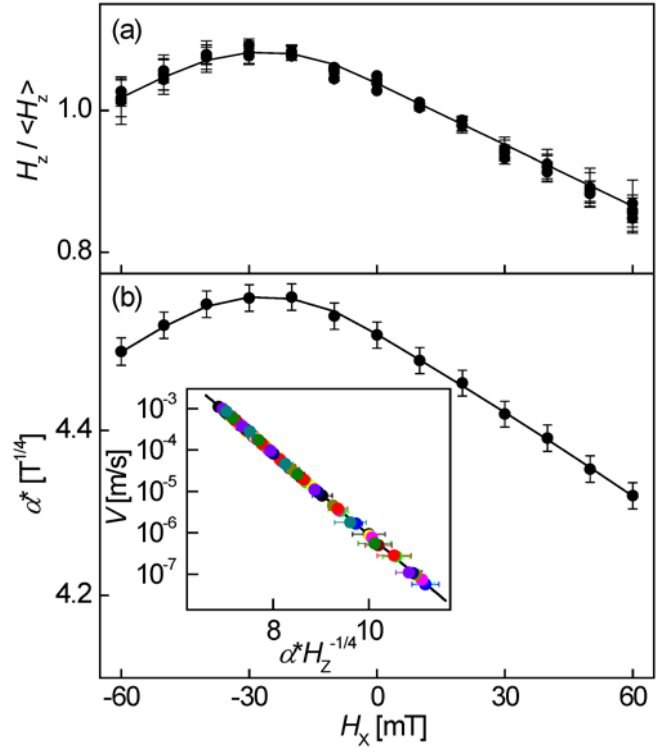

**FIG. 3.** (a) $H_z/\langle H_z\rangle$ and (b) $\alpha^*$ with respect to $H_x$. The solid lines show the best fit using Eq. (6). The inset shows $V$ vs. $\alpha^* H_z^{-\mu}$ for all experimental data. Each color of the symbols corresponds to a different $H_x$. The solid line shows the best linear fit.

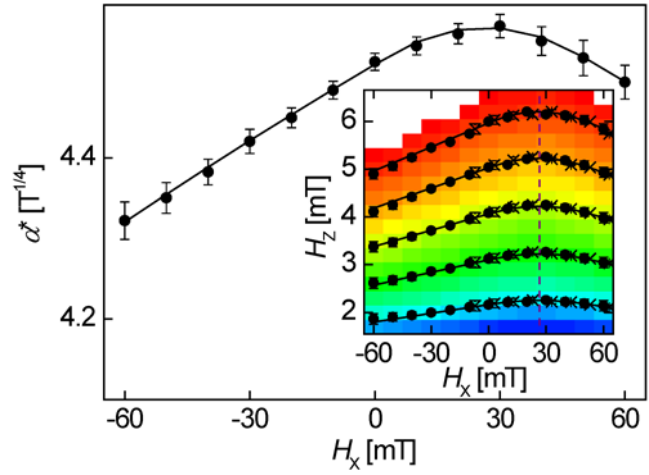

**FIG. 4.** $\alpha^*$ vs. $H_x$ of the latter experiment with opposite magnetic polarities. The solid line is obtained using the best fitting values for Fig. 3, except the opposite polarity of $H_{DMI}$. The inset shows the equi-speed contour map. The symbols and the lines are identical to those in Fig. 2.